# QoS-enabled ANFIS Dead Reckoning Algorithm for Distributed Interactive Simulation

Akram HAKIRI [1, 2], Pascal BERTHOU[1, 2], Thierry GAYRAUD[1,2]

[1]CNRS ; LAAS, 7, avenue du Colonel Roche, 31077 Toulouse, France

[2] Université Toulouse; UPS, INSA, INP, ISAE; LAAS; F-31077 Toulouse, France

Email: {Hakiri, Berthou, Gayraud}@laas.fr

**ABSTRACT**: *Dead Reckoning mechanisms are usually used to estimate the position of simulated entity in virtual environment. However, this technique often ignores available contextual information that may be influential to the state of an entity, sacrificing remote predictive accuracy in favor of low computational complexity. A novel extension of Dead Reckoning is suggested in this paper to increase the network availability and fulfill the required Quality of Service in large scale distributed simulation application. The proposed algorithm is referred to as ANFIS Dead Reckoning, which stands for Adaptive Neuro-based Fuzzy Inference System Dead Reckoning is based on a fuzzy inference system which is trained by the learning algorithm derived from the neuronal networks and fuzzy inference theory. The proposed mechanism takes its based on the optimization approach to calculate the error threshold violation in networking games. Our model shows it primary benefits especially in the decision making of the behavior of simulated entities and preserving the consistence of the simulation.*

## 1. Introduction

Distributed Simulation applications require decision making due to the distributed nature of the simulated entities. Updating these entities states generates a high quantity of information to be sent over physical link. Moreover, the Quality of Service (QoS) and the CPU processing require theoretical as well as experimental investigation in order to optimize and to better help the management of the available network resources. Dead Reckoning (DR) algorithms were proposed as message filtering techniques. This technique consisting in the estimation one's current position of an entity based upon a previously determined position and advance this position based upon known or estimated speed over elapsed time. Many applications involve that algorithm using a single linear or quadratic extrapolation model in order to preserve the consistency among players and the causality for networked games.

The aim of this study is to focus on the error threshold violation induced by the DR mechanism. This paper is twofold: 1) suggest an investigation of the Quality of Service (QoS) requirements in large scale distributed simulation applications, 2) explore the feasibility of using DR system based on an artificial intelligence model and attempt its ability to address the intelligent predictive DR mechanism. This last contribution also is double: a) provide a mathematical formalism to found an admissible threshold error to satisfy the QoS requirements from the users and the network viewpoint, and endow with an optimization technique to choose an optimal value of this error, b) use an adaptive technique to adapt the error with context of the simulation based on Adaptive Neuro-Fuzzy Inference System (ANFIS) in order to refine the optimized error.

This paper is organized as follow: after a brief introduction in Section 1, an overview of the Dead Reckoning is given in Section 2. Section 3 presents some researchers opportunities which attempt to suggest prediction model to the DR process. The reminder of Section 4 is to provide generalized QoS formal model to support the user behavior and make available QoS specification which captures application QoS requirements. In Section 5, we point on the method that we used to fulfill these QoS constrains using formal mathematical approach. Section 6 introduces ANFIS Reckoning model that can adapt the threshold error violation to generalized context of the distributed interactive simulation. Conclusion is given in section 7.

## 2. Dead Reckoning Algorithm

The IEEE1278 protocol [1] provides a standard set of 9 algorithms for entity position and orientation dead reckoning [23]. This technique dates back to the navigational techniques used to estimate ship's current position based on start position, travelling velocity and elapsed time. DR is used to reduce the bandwidth consumption by sending update message less frequently due the error threshold violation and estimating the state information between updates.

Each remote site maintains in addition to the high fidelity model of the position and the orientation of the entity it generates a dead reckoning model to estimate the entities (whether they are computer generated or human-in-the-loop simulator) behaviors used when these differ by a threshold. For simulated entities, the process is to assign an algorithm to a given entity and continue using dead reckoning during the simulation. When the gap between the extrapolated states and the real states exceeds the defined threshold ($TH_{pos}$ for the position and TH$_{or}$ for the orientation), the simulator transmits messages more frequently to anticipate the entities states motion. The anticipated entity states are computed from the last states based on the extrapolation of the position; the velocity, the orientation and the acceleration of the entity.

On all other remote sites, the reception of new packets implies the updating of the entity states following some standard trajectories like circular, elliptical, ballistic, parabolic, and hyperbolic path [23]. Thus, often the first order or second order (equation 1)) extrapolation equations are used:
- Second order extrapolation: in addition to the given $P_i$ and $V_i$ from the first order extrapolation, the acceleration $A_i$ is added to the second order extrapolation. So, the quadratic extrapolation is given by :

$$P_{DR}(t) = P_i + V_i(t - t_i) + A(t - t_i)2 \qquad (1)$$

### 3. Related work

The DIS protocol [1] does not define how to calculate the error. Consequently, there has been a tremendous amount of feedbacks from the users of the distributed simulation applications regarding the DR tests. Amaze [26] represents one of the earliest multiplayer games which implement the dead reckoning technique in distributed simulation. Authors in [2] present a dynamic filtering technique in order to send messages and introduce a dynamic threshold denoted by Update Lifetime (UL). UL is the delay between two consecutive updates and it aims to make a new dynamic DR threshold. If UL is smaller, then an important threshold is produced to increase the performance of the communication. Otherwise, a small Threshold will be produced.

Authors in [3] introduce also a technique referred to Variable Threshold for Orientation (VTO). An auto-adaptive DR algorithm was presented in [4]. It fuses a dynamic multi-level DR threshold with the relevance filtering. It uses the distance between two simulated objects to calculate the threshold: a small distance involves a small threshold whereas a larger distance induces more important filtering of messages, which ensure the scalability of the simulation.

The position history-based protocol [25] is dead reckoning approach which transmits timestamps packets containing the entity state. It maintains a history of position and uses curve-fitted technique to predict future entity positions.

Two other adaptive algorithms are described in [5]. The first is centered on an adaptive adjustment of the threshold, whereas the second benefits from the previously determined values of the threshold to automatically select the extrapolation equation. However, it needs always data collected from the live simulation. Authors in [6] present a deterministic estimator of the objects parameters (position, velocity…). Nevertheless, this model will generate a larger estimated error when an unpredictable behavior of the entities occurs. Kalman filters were also present in [7] to estimate the mobility of ad-hoc mobile nodes and then reduce the unwanted traffic and it was also suggested for humanoid robot motion [8].

A fuzzy logic approach based on the fuzzy correlation degree of all measured parameters of the entities (position, size, vision angle, velocity…) was proposed in [9], and it considers a multi-level threshold on which applications should obey. A Nero Reckoning algorithm was suggested [10, 13 and 14] as an intelligent approach to resolve the complexity of some sophisticated polynomial extrapolation. The remote host uses a neuronal banc that includes the desired parameters (position, the velocity, the orientation…) to produce the new predicted proprieties of the simulated objects.

### 4. QoS requirements in distributed simulation

An application's QoS requirements are conveyed in terms of high-level parameters that specify what the user requires. QoS specification encompasses requirements for distributed simulation to characterize the degree of:

- Coherence: the coherence of the simulation consist of the spatial coherence and the temporal coherence:
  - The spatial coherence requires that at any time of the simulation, the gap between the entity state in the sender site $S_e$ and that in the receiver site $S_r$ does not exceed the threshold. For example, in Figure 1, the gap of the position shift represented by $E_p$ should fulfill the following conditions: $Th_{pos} \geq |E_{pos}|$ and $Th_{or} \geq |E_{or}|$.
  - The temporal coherence involves that every remote site have to know the occurrence of all events happening on all other sites within a bounded delay.

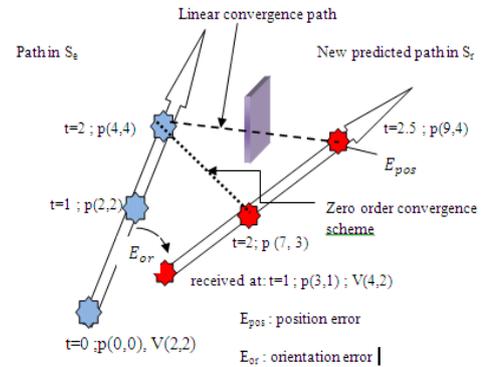

Figure 1: Dead Reckoning with prediction and convergence scheme

- Performance: expected performance characteristics are needed to establish resource commitments, and can be expressed using three terms:
  - Reliability: the maximum allowable error packet lost, denoted $\tau$, which is closely related to the maximum admissible error to ensure the spatial coherence of the simulation.
  - Latency: maximum allowable delay across the network denoted DT, which is directly related to the temporal coherence.
  - Jitter: the temporal deviation of this latency referred to as $\Delta DT$.

- Coupling: the QoS parameters are specified in accordance with the coupling level [11] between entities:
  - Tightly coupled interactions: this type of coupling occurs when several entities move within a narrow zone. So, transmitted data packets require more performances to ensure the coherence and the consistency of the simulation. In such case, the latency is defined as $DT \leq 100$ ms and the error packet lost is given as $\tau \leq 2\%$.
  - Loosely coupled interactions: this kind of coupling happens when entities are fairly numerous and the distance which separates them is large enough to tolerate the transmission errors. Latency requirements between the output of data packet at the application level of a simulator and input of that data packet at the application level of any other simulator in that exercise is defined as $DT \leq 300ms$, and the error packet lost is given as $\tau \leq 5\%$.

Indeed, to provide and sustain QoS, resource management must be QoS-driven. Hence, expressing QoS regarding the coupling level has three major drawbacks:
- L1: the determination of the coupling level between distributed entities (which may evolve over time) seems to be expensive in terms of computation time.
- L2: the expression of the QoS based on the coupling degree (sender sites and receiver sites) sits uneasily with the multicast transport [28].
- L3: the distributed simulation standards [1, 16] ignore the influence of the network latency on the position/orientation errors in large scale distributed simulation applications, and neglect the spatial coherence constrains.

From the inspection of these drawbacks, we investigated the influence of the threshold error violation on the QoS required at the user level. Hence, let's consider Figure 2, and assuming that both the sender site Se and the receiver Sr are implicated in the same simulation exercise and exchange data packets including the position, the velocity and the acceleration of the entity denoted A. We focus on the visible behavior of A, and assuming that the DR approach is applied to reduce the network traffic. Figure 2 illustrates the extrapolated error of the position of the entity A in both $S_e$ and $S_r$ sites, that we note respectively $E_s$ and $E_r$.

We chose the black circles to illustrate the packets sent from the sender site Se and the gray circles to indicate the reception of these packets at the receiver site Sr.

Dates Te0, Te1, Te2 and Te3 illustrate the date when packets are sent from site Se, and respectively dates Tr0, Tr1, Tr2 and Tr3 indicate the date when packets are received on site Sr.

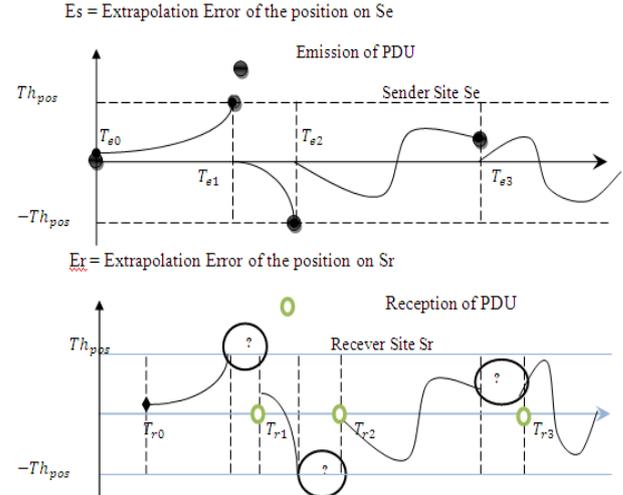

Figure 2: Evolution of the extrapolated error on both the sender site Se and the receiver site Sr.

Let's examine the error at the both the sender site Se and the receiver site Sr:

On the sender site $S_e$:
- At $T_{e0}$, the extrapolation error of the position of the entity *A* increases and reaches its maximum value (Thpos or DR) at Te1: at that time the DR model begins sending data packets to correct the path with the right value of the position of A and the error Er become near to zero.
- From the time Te1 the previous scenario is reproduced until the time Te2.
- At time Te2, the threshold error oscillates between Thpos and –Thpos without leaving this interval, during the HEART BEAT TIMER, that to say Te3 = Te2 + 5s. The DR model in the sender site Se transmits packets to bring error to zero.

On the receiver site $S_r$:
- At time $T_{r0}$, the extrapolation error Er remains the same as in site Se at time Te0; particularly, the error $E_r$ reaches the maximum allowed value (Thpos at Te1, the date when refresh packets are sent). However, the update of the position of A is made at time Tr1. The interval between [Te1, Tr1] corresponds to the network delay, where packets move from the sender to the receiver. In fact, it seems within this interval the error $E_r$ is unpredictable and can exceed the Dead Reckoning threshold (Thpos) generating the spatial incoherence.
- We highlight the lack of control of the error $E_r$ during the interval marked in question mark.

From the inspection of these analyses, it seems that the spatial coherence during the simulation is not guaranteed, except in the interval [Te0, Tr0] separating the emission and the reception of update packets; during other periods the error Er exceed (the absolute value of $E_r$) transiently the maximum allowed value.

Thus, two questions can be asked: (1) does the spatial incoherence is prejudicial to the progress of a correct simulation exercise? (2) If so, how can we resolve this problem in order to control the transiently exceed of the error Er?

Concerning the first question, the transiently exceed of the error Er becomes detrimental when the latency (DT) is important ahead the period separating the reception of two consecutive refresh packets. Subsequently, the remote sites will have an incoherent spatial view during several times. This risk appears in large scale distributed networks.

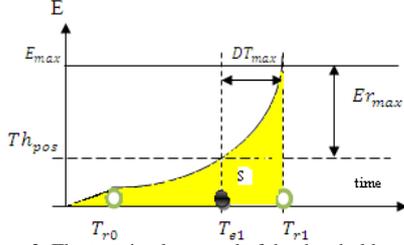
Figure 3: The transiently exceed of the threshold error Er

To answer to question (2), we need a new investigation to fulfill the previous illustrated QoS requirements. Therefore, in order to guarantee that the threshold error violation $E_r$ does not exceed the maximum allowed value, We need to found this maximum admissible error that we note $E_{max}$ during the delay $DT_{max}$ (see Figure 3).

In fact, the excess of the error is directly related to the dynamical behavior of the entity, as consequence the maximal allowed error $E_{rmax}$ is closely related to the maximum network delay $DT_{max}$ as shown in Figure 3.

Note that this threshold error is the worst case, because it aims to find a limit "M" to the function characterizing the trajectory of the simulated entity. So, the function $r(u,t)$ describing the entity motion has a limit $M$ at an input close to the $\infty$ if this function $r$ is close to M whenever t is close to ∞. In Section 5, we will provide a method to found the excepted limit "M", and then we will use an optimization method to refine the threshold error far from the worst case.

## 5. Mathematical formulation of the error

In this Section, Our contribution is twofold: 1) we suggest a formal mathematical approach to find the worst case admissible extrapolated error and estimate its value from the behavior of the entity, and 2) we look into an optimization method to refine this worst case, to be injected on an adaptive mechanism based on the Neuro-fuzzy reasoning to build a DR model that imitate the high fidelity model.

We consider a virtual infinitesimal displacement of the entity which refers to change in the trajectory of this entity as the result of the change of any infinitesimal change of the coordinates *δu* and *δt*, consistent with the constrained (error threshold violation) imposed on the entity at any given time *t*. This displacement is called virtual and denoted (δ) to distinguish it from the actual displacement of the entity, denoted (*d*) occurring in a time interval *dt*, during which the constrains may be changing. We can obtain the entire motion of the entity between times *δt1* and *δt2* and we can get the small variation of the entity motion from the actual motion. For simplicity, we are going to look at one-dimension, although these results works just fine for d-dimension as well.

### 5.1. Characterization of the threshold error

Suppose that we have a nice surface S and a function $f: S \to \Re$ defined on the surface. We want to define an integral of $f$ on S as the limit of some sort of Riemann sum. To simplify the presentation, we assume the surface is sufficiently smooth to allow us to approximate the area of small piece of it by a small planar region, and then add up these approximations to get a Riemann sum. One method to calculate the surface integral is to split the surface to several small pieces $S1, S2, \ldots, Sn$ each having small area $\delta S$, then select points $r_i = (u_i, t_i) \in Si$, and finally form the Riemann sum. We take finer and finer subdivisions and the Riemann sums have a limit, and we call this limit the integral of $f$ on S (c.f relation (2)):

Note that to find an explicit formula of this integral, we need to parameterize S by considering on S a system of curvilinear coordinates, and for example the parameterization be r(u, t), where (u, t) varies in some region D in the plane (see Figure 4). Then, we need a vector description of S, say $r: D \to r(D) = S$. The surface is subdivided by subdividing the region $D \subset R^2$ into rectangles. Let's look closely at one of the subdivisions:

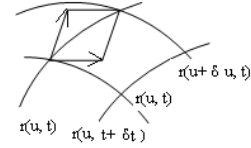
Figure 4: one subdivision area of S

We paste a parallelogram at the point $r(u_i, t_i)$ as shown in Figure 4. We calculated the lengths of the sides of this parallelogram in order to get some sort of the Riemann sum given in equation (2). This sum reaches the limit of the integral across the surface S when the variable u and t are close to∞:

$$R = \sum_{i=0}^{n} f(r(u_i, t_i)) \times \left|\left(\frac{\partial r}{\partial u}(u_i, t_i)\right)\left(\frac{\partial r}{\partial t}(u_i, t_i)\right)\right| \partial u \partial t \quad (2)$$

In distributed simulation applications, the surface S is the surface that we can get between the Dead Reckoning path and the real straight path when the entity moves during the simulation. In fact, the Riemann sum is a discretization of the general function given in the equation (3). Using this approximation, we can found a limit "M" that we expect and help us to find the distance between the real position (vector) on which the entity is supposed to be on and the position (also vector) of the entity when the DR mechanism is applied. So, within the plan D we can get the formula:

$$R' = \iint_S f(r)dS$$
$$= \iint_D f(r(u_i t_i)) \left|\left(\frac{\partial r}{\partial u}(u_i, t_i)\right)\left(\frac{\partial r}{\partial t}(u_i, t_i)\right)\right| dS \quad (3)$$

Let's use our new-found knowledge and the extrapolation equation founded in the equation (1) to calculate the threshold error violation. Note that the effective position curve is denoted by $P_a(u, t)$, the extrapolated motion curve is denoted $P_{DR}(u, t)$ and the gap between the effective curve and the DR curve at any time is denoted $E_p(t)$. Using the triangle inequality in Normed Vector Space V (equation (4)) we find the limit of the extrapolated error.

$$\|x + y\| \leq \|x\| + \|y\| \; \forall \, x, y \in V \tag{4}$$

From the above discussion, we note the existence of real number $E_{max}$ which implies the following:

$$\begin{aligned} E_p(t) &= \|P_a(t) - P_{DR}(t)\| \\ &= \left\| \int_{T_{ei}}^{u} du \int_{T_{ei}}^{u} [A_a(t) - A_i(t)] d\tau \right\| \\ &\leq \int_u du \int_{T_{ei}}^{u} |A_a(\tau) - A_i| d\tau \; + \\ &\quad \int_u \int_{T_{ei}}^{T_{ei+1}} [A_a(t) - A_i(t)] d\tau \; + \\ &\quad \int_{T_{ei+1}}^{T_{ei+1}+DT} \int_{T_{ei+1}}^{u} [A_a(t) - A_i(t)] d\tau \\ &\leq E_{max} \end{aligned} \tag{5}$$

Note $A_a(t)$ and $A_i(t)$ are respectively the effective instantaneous acceleration of the simulated entity at any time (t) and its acceleration at time $T_{ei}$. Note also that the error can be approximated to the Euclidean distance D (see equation 6). So, we can calculate this distance using three terms in equation (5): (1) the DR Threshold (Thpos in our case, just to simplify) (2) the norm of the acceleration (considering any acceleration vector in the curve, we can calculate it by assuming a maximum value which we note $A_{max}$) and (3) the term $(V(t) - V(T_{ei+1})) \times DT$.

Thus, if we can fix the Threshold value and we can found the norm of the acceleration vector, we will need just to obtain a value of the third term. Indeed, using equation (3) we found the expected worst case limit of the threshold error which is the maximum admissible value of the threshold error violation. This value should not be exceeded in order to preserve the minimal coherence in the distributed simulation exercise, i.e. if we exceed this limit the spatial incoherence which is prejudicial to the progress of a correct simulation exercise will be not respected.

In addition, we need to refine the error using an optimization approach.

We used the variational form of the Euclidean distance between two objects $A_a$ and $A_b$ defined by $A_i = \vec{r}(u,t)$ and $A_a = \vec{r}(v,t)$. The object $A_a$ is located at the extrapolated curve and the object $A_i$ is located at the motion curve of the entity at any time, and we get something like this:

$$D = \int_u^t \int_v^t dudt \left\{ \sqrt{\left(\frac{\partial \vec{r}(u,t)}{\partial t}\right)^2} + \lambda \left[ \sqrt{\left(\frac{\partial \vec{r}(u,t)}{\partial t}\right)^2} - 1 \right] \right\} \tag{6}$$

$\lambda$ is the Lagrange multiplier. We need is to choose the best value of the Lagrange multiplier $\lambda$ which provides a way to resolve our optimization problem; technically, the Lagrange multiplier corresponds to the point where the differential of the function $f(u,t)$ has an orthogonal kernel to the gradient to the subject function $g(u,t)$ in this point (the principle of the resolution of the Lagrange Equation).

Since we founded an expression to distance between two objects respectively on the high fidelity path and DR path, we will proceed to the optimization of this error.

### 5.2. Optimization technique

Let's consider the same case presented in Section 4 which we reproduce in Figure 5. We have a function f (u, u', t) defined on the path of the simulated entity (we consider the function u (t) depending on the time t between times δt1 and δt2. We are going the find the particular path u (t) such that the line integral defined by the equation (7)

$$J = \int_{\delta t1}^{\delta t2} f(u, u', t) dt \tag{7}$$

has a stationary value relative to the path differing from the correct function $u(t)$. This will allow as using such a magnifier at each infinitesimal point of the entity trajectory. We call the relation (7) the optimization criteria (Hamilton principle).

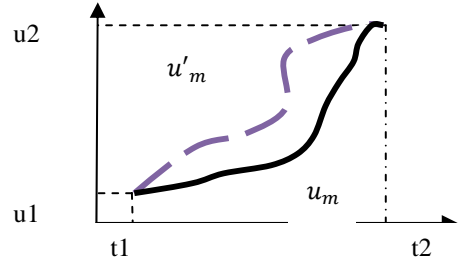

Figure 5: approximation approach

$$u'_m = u_m + \lambda n(t) \tag{8}$$

Let's $u_m$ the real path that the entity should follow and $u'_m$ the path generated by the threshold violation, where $\lambda$ is an infinitesimal parameter and n (t) any function that vanishes at t1 and t2.

For any curve in Figure 5 the integral J (equation (8)) is also function of $\lambda$:

$$J(\lambda) = \int_{\delta t1}^{\delta t2} f(u(t, \lambda), u'(\lambda, t), t) dt \tag{9}$$

Since J has a stationary value for the high fidelity trajectory (corresponding to $\lambda = 0$), we have:

$$\frac{dJ}{d\lambda}\bigg|_{\lambda=0} = 0 \tag{10}$$

Differentiating $J(\lambda)$ we get:

$$\frac{dJ}{d\lambda} = \int_{t1}^{t2} \left( \frac{\partial f}{\partial u} \frac{\partial u}{\partial \lambda} + \frac{\partial f}{\partial u'} \frac{\partial u'}{\partial \lambda} \right) dt \tag{11}$$

For α=0, the above equation gives the equation (12):

$$\int_{t1}^{t2} \left[ \frac{\partial f}{\partial u} - \frac{d}{dt} \left( \frac{\partial f}{\partial u'} \right) \right]_{\lambda=0} \tag{12}$$

So, the function $u_m$ for which the line integral J has a stationary value solves the Lagrange equation (13):

$$\frac{\partial f}{\partial u} - \frac{d}{dt} \left( \frac{\partial f}{\partial u'} \right) = 0 \tag{13}$$

The assertion that J is stationary for $u_m$ can be also written as:

$$\partial J = \partial \int_{\delta t1}^{\delta t2} f(u(t, \lambda), u'(\lambda, t), t) dt = 0 \tag{14}$$

Note that the above equation can be easily generalized to the case where $f$ is a function of many independent parameters $u_i$ and their derivatives $u'_i$ (for example when the entity is formed by several articulations and we need to calculate the error threshold for each articulation, see equation 15):

$$J = \int_{\delta t1}^{\delta t2} f(u_1, \ldots, u_n, u'_1, \ldots, u'_n, t)dt \quad (15)$$

Form the inspection of equations (12-15), we can see that the error threshold violation can be optimized without using a multi-level threshold. If the constrain criteria J remains equal to zero, then the error threshold is also equal to zero and the ideal case will happen without threshold violation.

Further, if the curve $u'_m$ tends to be equal to $u_m$ then we can provide the end-to-end QoS guarantees in terms of bandwidth and network latency. This means that the distance (equation 6) can be approximated to zero when J is equal to zero. Note that the Lagrange multiplier in equation 6 is just near to zero but, it cannot be zero and also the same case for the constrain criteria J which tends to zero without being null.

In addition, the error threshold id closely related to the frequency of update messages $f_s$ [20].

$$f_s = \sqrt[3]{\frac{\ddot{u}}{3!Ep}} \quad (16)$$

Especially for a circular motion, the update frequency is given by the equation (16):

$$f_s = \sqrt[3]{\frac{Aa^2}{6VE_p}} \quad (17)$$

$A_a$ is the effective acceleration, $V$ is the velocity and $E_p$ is the threshold error. In fact, when the threshold error is minimal, the update frequency decreases proportionally. The equations of motion are given as:

$$\begin{cases} u_x = r.\cos\omega t; \; u_y = r.\sin\omega t \\ \dot{u}_x = -r\omega.\sin\omega; \; \dot{u}_y = r.\omega.\cos\omega \\ \ddot{u}_x = -r\omega^2.\cos\omega; \; \ddot{u}_y = -r\omega^2.\sin\omega \end{cases} \quad (18)$$

From this discussion, we can found a value of both the velocity and the acceleration and we can get a fine value of the threshold error which optimizes the update frequency. Then, we proceed to an intelligent technique that can predict an optimal value of the error. Consequently, both the network resource and CPU usage can be optimized. Bandwidth and latency can be managed by the usage of our approach.

## 6. Adaptive Neuro-Fuzzy Inference System based Dead Reckoning

In order to optimize the new-founded value of $Ep$, we will proceed to an artificial intelligent mechanism to minimize the maximum allowed error which satisfies the spatial and the temporal coherence of the simulation.

Our goal is to imitate the high fidelity model using the fuzzy correlation distance [9] between the real path and its extrapolated path of the entity. The distance is calculated using the general Euclidian distance (see equation 6) between two points virtually located on the extrapolated path and the high fidelity path after its refinement using the optimization approach (see equation 11), and then the refined distance is injected within an adaptive system to optimize the error threshold violation.

Our approach to fulfill the management of the QoS requirement in distributed simulation applications involves replacing the Dead Reckoning predictors by Neuro-fuzzy model that learn from the QoS required by the user experience during the simulation. The ANFIS [12] which stands for Adaptive Neuro-based Fuzzy Inference System is a technique based on fuzzy inference system trained by the learning algorithm derived from the neuronal network theory.

In order to evaluate the algorithm, experiments were conducted from the perspective of the controlling host. A simulation program was created to simulate the entity motion. The threshold errors were collected during the simulation. An entity moves in site Se in sinusoidal curve and trace file was used to collect statistics.

### 6.1. The ANFIS Reckoning Model

The fuzzy reasoning concerns deciding future actions on the basis of vague and uncertain knowledge of the previous information of the entity.

Fuzzy reasoning:
In a fuzzy set, the transition from "belonging to a set" and not "belonging to set" is gradual and this smooth transition is characterized by membership functions that give the fuzzy sets more flexibility in modeling the linguistic expressions.

Definition 1:
If U is collection of objects (called universe of discourse) denoted by *x,* then the fuzzy set A in U is defined as a set of ordered pairs:
$$A = \{(x, \mu_A(x)) | x \in U\} \quad (19)$$
$\mu_A(x)$ is called the membership function (MF) of x in A. The MF maps each element of U to continuous membership value between 0 and 1. We shall note the existence of several classes of parameterized functions commonly used to define MF's. We used in our approach, the generalized bell function (see Figure 6).

Definition 2:
Let's U a subset of $\Re^n$. The distance transform (DT) of S is given as an image $\{(x, D_s(x)) | x \in \Re^n\}$ on $\Re^n$ where $D_s$ is the DT value on $x$ that is defined as follow:
$$D_s(x, y) = \inf\{\|x - y\|\| x, y \in \Re^n\} \quad (20)$$
Where, *inf* gives the infimum of a set of positive numbers and $\|.\|$ is the Euclidean norm.

This expression of the fuzzy distance is generalized to the Euclidian distance for objects given in equation (6). When considering two objects A and B, the ANFIS threshold is used to determine whether send the update packets. Equation (6) and its refined version in equation (11) are taken into account to calculate the extrapolated error.

Our ANFIS Reckoning Model is composed of three inputs (objects) denoted (a1, a2, and a3) which refer to the current position, the velocity and the acceleration of the entity and only one output the error threshold. The mapping output function f is given by:
$$f^k = f(x^k) = f(a_1^k, a_2^k, \ldots, a_n^k); k \in 1..K \quad (21)$$
The learning process associates the three inputs, at any time, to the output in order to converge to the finally estimated output as shown in equation (22)
$$\{(a1^1 a2^1 a3^1 f^1), \ldots, (a1^k a2^k a3^k f^k)\} \quad (22)$$

The position, the velocity and the orientation properties were employed in the simulation. The orientation includes the view direction in the interval $\left[-\frac{\pi}{2}, \frac{\pi}{2}\right]$, the position was chosen regarding the fixed threshold in the interval $[-Th_{pos}, Th_{pos}]$ and also the velocity interval is given in the interval $[-V_{max}, V_{max}]$. The acceleration vector is calculated based on the velocity vector.

It should be noted that our objective is to make more refinement to the threshold error regarding three terms: the threshold, the acceleration and the velocity and we need to estimate the new updated position with the respect of the admissible error. Indeed, each criteria associated to the problem is fuzzified by defining a membership function that correspond to the Sugeno [16] "IF-THEN" rules behind the criteria.

Furthermore, in our approach, we need membership function to describe the different attributes (inputs) using linguistic variables and every linguistic variable can have seven linguistic terms shown in Figure 7:{NB, NM, NS, ZE, PS, PM, PB} (the training sets derived from this terms can be written in the form Negative Big, Negative Medium, Negative Small, Zero, Positive Small, positive Medium, Positive Big*)* and their membership function are of the sigmoid form (see equation 23) characterized by three parameters:

$$sigmoid(x; a, b, c) = \frac{1}{1+exp\,[-a(x-c)]} \quad (23)$$

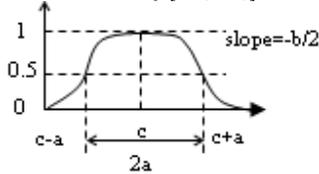

Figure 6: Meaning of parameters in generalized bell function

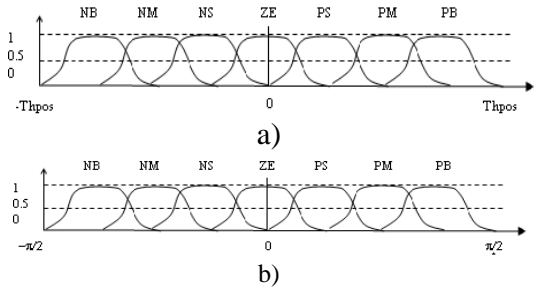

Figure 7: linguistic terms for a) the position and the velocity variables, b) for the direction variables

We used the following rules (note that we need 343 rules to fill all the criteria, but using our tested we only used 16 rules, and we just illustrate 3 of them):

ℜ1: *if* a1 is PS and a2 is PM and a3 is PM *Then* O is α1
ℜ2: *if* a1 is PB and a2 is PB and a3 is PB *Then* O is α2
ℜ3: *if* a1 is PS and a2 is PM and a3 is PM *Then* O is α3

Figure 8 illustrate the output of each fuzzy rule. The parameters $\alpha_i$ vary during the simulation in the first of the ANFIS reckoning layer (see Figure 9). Using this approach we can refine more and more the extrapolated error to get the best value before sending it to the final output.

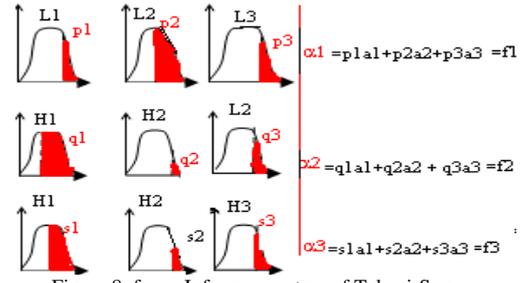

Figure 8: fuzzy Inference system of Takagi-Sugeno

For example, when the position of the entity is in close proximity to *Thpos* the value of the position is taken from the membership function of each curve within its definition interval, and so on. These values are not fixed by the user, but the ANFIS algorithm calculates (adjusts) them during the simulation and applies the value needed at each step of the simulation.

Indeed, the ANFIS algorithm estimates continually the information (position, direction, velocity…) sent from each remote simulator and at each step it can give an idea about the behavior of the simulated entity. From the inspection of Figure 8, the vague and unclear values of the position make the problem of decision making under uncertainly, especially with the information we have about the possible outcomes of the outputs, the value of the new updated information, and the dynamically changing condition is vague and ambiguous. Note that each membership function of each input is adjusted with the respect of the output of the ANFIS reckoning model.

Indeed, when defining fuzzy sets, we inevitably face the question, how should we assign the membership functions. Here, we suggested the adaptive method: the membership functions are dynamics and evolve over time using the feedback from the input data of the neuronal network banc. So, there are an infinite number of possible different membership functions (with variables values of parameters in equation 23) for the same attribute, and by tweaking these membership functions we can get more accurate response.

The ANFIS network used in this investigation was a five layer networks (Figure 10), with six nodes in the first layer representing the each dimension of the input vector, one node in the last layer representing the output, and 3 hidden layers consisting of three nodes in each layer. This network attempts to develop a matching function between the input and the output vectors by using some training algorithms.

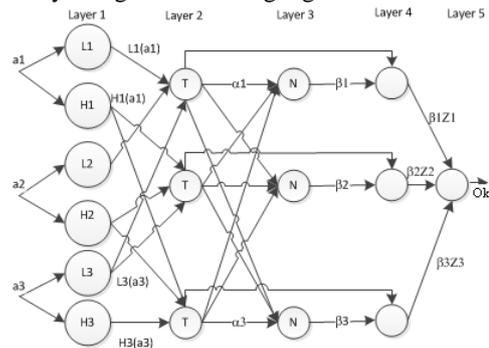

Figure 9: ANFIS Architecture of the DR Algorithm

Figure 8 describes the ANFIS architecture of the DR Algorithm. Each layer in this architecture has specific goal:

**Layer 1:** the output of each node is the degree to which the given input satisfies the linguistic term associated to this node. The values of the fuzzy members are calculated in this layer.

**Layer 2:** each node realizes a T-norm function (many other function can be chosen depending on the goal of the ANFIS approach) to compute the firing rules. The outputs of these nodes (called rule nodes) are given by something like this given in relation (24)

$$\begin{cases} \alpha_1 = p_1 a_1 + p_2 a_2 + p_3 a_3 \\ \alpha_2 = q_1 a_1 + q_2 a_2 + q_3 a_3 \\ \alpha_3 = s_1 a_1 + s_2 a_2 + s_3 a_3 \end{cases} \quad (24)$$

**Layer 3:** the outputs of the T-norm are normalized in this layer. The output f the top, middle and bottom neuron is the normalized firing level of the corresponding rule:

$$\begin{cases} \beta_1 = \frac{\alpha_1}{\alpha_1 + \alpha_2 + \alpha_3} \\ \beta_2 = \frac{\alpha_2}{\alpha_1 + \alpha_2 + \alpha_3} \\ \beta_3 = \frac{\alpha_3}{\alpha_1 + \alpha_2 + \alpha_3} \end{cases} \quad (25)$$

**Layer 4:** the learning process is done in this layer. The instantaneously calculated output is compared to the estimated output. If the off line learning technique is used, the estimated output is stored in database and then compared the output of the algorithm. The output of each neuron is given by the normalized product of the firing rule and the output correspondent rule, like shown in relation (26):

$$\begin{cases} \beta_1 z_1 = \beta_1 \times \alpha_1 \\ \beta_2 z_2 = \beta_2 \times \alpha_2 \\ \beta_3 z_3 = \beta_3 \times \alpha_3 \end{cases} \quad (26)$$

**Layer 5:** this is a single node of the output function of the overall system, as the sum of all incoming signals.

$$O_k = \beta_1 z_1 + \beta_2 z_2 + \beta_3 z_3 \quad (27)$$

Rather than choosing the parameters associated with a given membership function arbitrarily, these parameters could be chosen so as tailor the membership functions to the input/output data in order to account for these types of variations in the data values. The Neuro-adaptive learning method works similarly to that of neural networks.

The ANFIS algorithm provides a method for fuzzy modeling procedure to learn information about a data set; it constructs a fuzzy inference system (FIS) whose membership function parameter are adjusted using either the back-propagation algorithm (gradient descent method in equation (28)) alone or in combination with the last square error type method (equation (29)), then it maps inputs through input membership functions and associated parameters, and through the output membership functions and associated parameters to outputs it can interpret the input/output map.

$$b_i(t+1) = b_i(t) - \eta \times \frac{\partial E}{\partial b_i} \quad (28)$$

$\eta$ is the learning rate of the gradient method and $b_i$ is the slope of our membership function associated with each input of the ANFIS model (see also equation 23 and Figure 6). Thus, the adjustment of the threshold error is made with the help of equation (29).

The parameters associated with the membership functions changes through the learning process. The computation of these parameters (adjustment) is facilitated by the gradient vector. This gradient vector provides a measure of how well the fuzzy inference system is modeling the input/output data for a given set of parameters. When the gradient is obtained, the optimization routine is applied in order to adjust the parameters to reduce some error measure. This is defined by the squared combination of the least squared error between the actual and the desired output given in the equation (29).

$$E_k = \frac{1}{2}(y_k - o_k) \quad (29)$$

Where $y_k$ is the $k^{th}$ component of the $p^{th}$ desired vector and $O_k$ is the $k^{th}$ component of the actual output vector produced by presenting the $p^{th}$ (Layer) input vector to the network.

From the relation (3 to8) we founded a way to calculate the surface as Riemann sum, as consequence we can easily found the error measure E for the $p^{th}$ of the training data as the sum of the squared error in all input layers:

$$E = \sum_{p=1}^{P} E_k \quad (26)$$

Therefore, we expect the choice of the best value of the admissible error. Obviously, when the error $E_p$ is equal to zero (after being adjusted using the first step of the refinement using equation (14) and then going through the ANFIS mechanism), the network can is able to bear all the traffic required for the consistency of the simulation and provides the required throughput. In that case, the ANFIS model succeeds in reducing both the spatial coherence and the temporal coherence associated with remote entity. Furthermore, typically an entity performs various movements in virtual environment, like circular, elliptic, ballistic, etc. the threshold calculated using the ANFIS Reckoning takes into account the possibilities of these motions and adapts the threshold regarding the considered motion at any time during the simulation.

Figure 10 shows the ANFIS logical diagram applied at each node: the high fidelity model stands for the model which the entity should follow. On the basis of this model, the low fidelity model is connected to the high fidelity model to check whenever the error threshold can reach its maximum allowed value using the segmentation (see section 5) approach, then the error is adjusted and refined using the Lagrange equation (see equations 12, 13 and 14) which take as function the path of the simulated entity.

After that, the ANFIS model is triggered and the ANFIS Dead Reckoning is applied. This process is applied in a loop until the output of the $5^{th}$ layer (the output error) is just accepted by the model.

Since we subdivided the region S into several infinitesimal parties, each one representing the local error calculated from the output of the ANFIS model, we used each subdivision to distribute the membership functions of the inputs of the Fuzzy Inference System (FIS) after the refinement of the error threshold. Thus, the selection of the membership value for each input is done by the gradient descent method. Therefore, each membership value is well located within the curve (see Figure 3).

In order to evaluate the performance of our ANFIS reckoning approach, we compared it with two other approaches: the position history based dead reckoning approach [25] is presented in Figure 11, and adaptive dead reckoning approach [5] is presented in Figure 12.

Since the position history used by remote tracking offers a fair estimate future position, it used a tight threshold that requires higher packet rates.

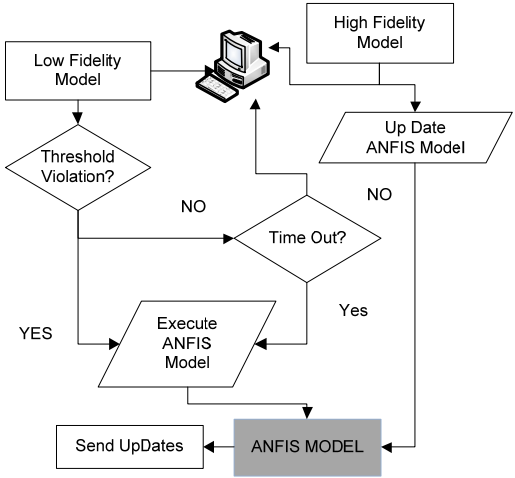

Figure 10: Proposed ANFIS Reckoning Model applied on each node

Further, the adaptive reckoning algorithm was used to control the accuracy of the extrapolation and influence the packet update frequency: a small threshold encourages more update packet generated more frequently, and a larger threshold value generates fewer packets sent at low update frequency.

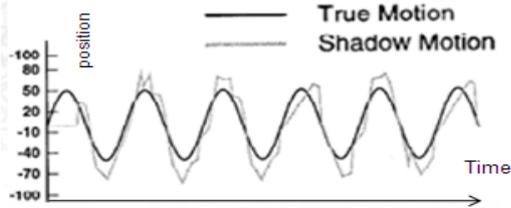

Figure 11: Extrapolation using the History based modeling of sinusoidal Oscillation [25]

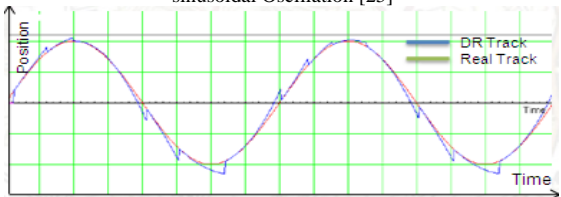

Figure 12: Extrapolation with Multi-level Threshold [5]

Figure 13 illustrates the trajectory of the entity when the threshold is adapted using the ANFIS Reckoning model. It can be seen that when entity is approaching, the extrapolated error becomes smaller and smaller. Therefore, the extrapolation of the trajectory at closed points to the real path is accurate to the real path to make the correct adjustment. Moreover, the update frequency is very low, so number of update packet is also very low. Consequently, by using an intelligent approach to adapt the fixed threshold without turn to multi-level threshold, we can fulfill the QoS required at the user level in terms of network latency and we can save considerably the bandwidth utilization as well as the low generation of update packets.

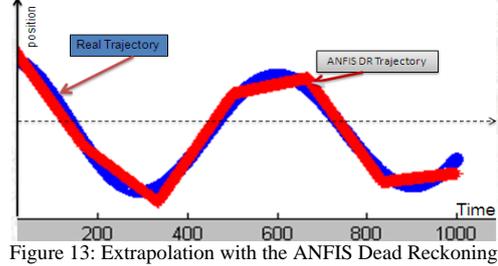

Figure 13: Extrapolation with the ANFIS Dead Reckoning

The results were compared to some other statistic results (see Table 1). The adaptive dead reckoning was used as predictive contract mechanism was presented [5]. From the inspection of the Table, we can note that ANFIS DR model has slightly better results to predict the best value of the threshold error. Additionally, the distance between the objects (see equation (6)) remains minimal when the Lagrange multiplier is very close to the gradient learning rate updated by the gradient descent method given in equation (28).

From these results, it is apparent that the neuronal network banc trained by the fuzzy inputs perform consistency better in tests regarding other approaches. Table 1 illustrates the threshold error performed with three approaches: the prediction of the threshold error using the ANFIS model yields an average error reduction up to 0.308 m. The two other results were also taken from paper [5], the threshold error is approximately 0.5608 m for the Area Of Interest (AOI) approach and 0.5573 for the Sensitive Region (SR).

Table 1: Threshold error comparison

|  | Threshold error (m) |
|---|---|
| ANFIS | 0.308 |
| AOI | 0.5608 |
| SR | 0.5573 |

We reproduced in Figure 14 after applying the ANFIS reckoning algorithm: the transiently exceed of the error to inspect the satisfaction the quality of service analyzed in Section 4. From these experiments, we can conclude that the ANFIS approach succeeded the reduction of the spatial error, and as a result the temporal error is also optimized to fulfill the QoS requirements from the users and application viewpoint.

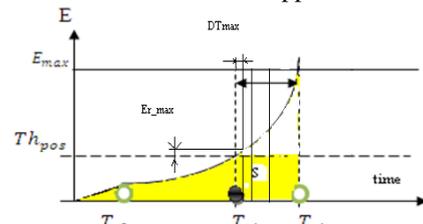

Figure 14: The transiently exceed of the error Er using the ANFIS reckoning approach

Moreover, the network latency and the bandwidth utilization are optimized by the new approach. The frequency of packet sent to update the trajectory of the simulation entity is optimized.

## 7. Conclusion

We presented in this paper an extension of the dead reckoning algorithm using hybrid learning algorithm based on Neuro-fuzzy technique, and we focused on the QoS specification of distributed simulation application in large scale networks. The proposed QoS-enabled Neuro-Fuzzy Dead Reckoning protocol provided more flexible scheme to fulfill the required QoS. It uses the benefits of empirical optimization technique to supply both the fuzzy inputs and the neural networks: Fuzzy logic can encode the threshold error directly using rules with linguistic labels, and then these quantitative labels are injected within the learning process of neural networks which automated this process using the back-propagation algorithm turned with the gradient descendant method and improved the performance of the developed algorithm. The overload of communication is lean and the bandwidth reducing is enhanced. Simulation results validate the potential of our proposed solution and explicit the efficiency of our decomposition scheme to improve the predictive performance.

Many more ideas, protocols and products have implemented some kind of DR algorithms. However, it is still a matter of research to find out which involve the QoS management over predictive protocols in distributed asynchronous communication networks.

## 8. Acknowledgment

This research is supported by the French FUI-DGE (Single Inter-Ministerial Fund of the Directorate General for Enterprise) program within the network simulation Platform (PLATSIM).